\documentclass[amsmath, amssymb, aps, prl, reprint, preprintnumbers, superscriptaddress, nofootinbib, nobibnotes]{revtex4-2}

\usepackage[usenames,dvipsnames]{color}
\usepackage[]{color}

\usepackage{graphicx}
\usepackage{lineno}
\usepackage{colortbl}
\usepackage{siunitx}
\usepackage{dcolumn}
\usepackage{blindtext}
\usepackage[caption=false]{subfig}
\usepackage[table, usenames, dvipsnames]{xcolor}
\usepackage{array}
\newcolumntype{L}[1]{>{\raggedright\let\newline\\\arraybackslash\hspace{0pt}}m{#1}}
\newcolumntype{C}[1]{>{\centering\let\newline\\\arraybackslash\hspace{0pt}}m{#1}}
\newcolumntype{R}[1]{>{\raggedleft\let\newline\\\arraybackslash\hspace{0pt}}m{#1}}

\newcommand{\triumf}{\affiliation{TRIUMF, 4004 Wesbrook Mall, Vancouver, British Columbia V6T 2A3, Canada}}
\newcommand{\nd}{\affiliation{Department of Physics, University of Notre Dame, Notre Dame, Indiana 46556, USA}}

\newcommand{\losalamos}{\affiliation{Theoretical Division, Los Alamos National Laboratory, Los Alamos, New Mexico 87545, USA}}
\newcommand{\cta}{\affiliation{Center for Theoretical Astrophysics, Los Alamos National Laboratory, Los Alamos, New Mexico 87545, USA}}
\newcommand{\ncstate}{\affiliation{Department of Physics, North Carolina State University, Raleigh, North Carolina 27695, USA}}

\newcommand{\xinst}{\affiliation{Key Laboratory of Particle Astrophysics, Institute of High Energy Physics, Chinese Academy of Sciences, Beijing, 100049, People’s Republic of China}}
\newcommand{\uvic}{\affiliation{Department of Physics and Astronomy, University of Victoria, Victoria, British Columbia, V8W 2Y2, Canada}}
\newcommand{\ubc}{\affiliation{Department of Physics and Astronomy, University of British Columbia, Vancouver, British Columbia, V6T 1Z1, Canada}}
\newcommand{\canpan}{\affiliation{CaNPAN, \href{http://www.canpan.ca}{http://canpan.ca}}}
\newcommand{\nugrid}{\affiliation{NuGrid collaboration, \href{http://www.nugridstars.org}{http://nugridstars.org}}}

\begin{document}
\title{Thallium-208: a beacon of \textit{in situ} neutron capture nucleosynthesis}

\author{Nicole~Vassh}
\email{nvassh@triumf.ca}\triumf
\author{Xilu Wang}
\email{wangxl@ihep.ac.cn}\xinst
\author{Maude Larivière}\triumf\ubc
\author{Trevor Sprouse}\losalamos\cta
\author{Matthew~R.~Mumpower}\losalamos\cta
\author{Rebecca~Surman}\nd 
\author{Zhenghai Liu}\ncstate
\author{Gail~C.~McLaughlin}\ncstate
\author{Pavel Denissenkov}\uvic\canpan\nugrid 
\author{Falk Herwig}\uvic\canpan\nugrid
\date{\today}

\begin{abstract}
We demonstrate that the well-known $2.6$ MeV gamma-ray emission line from thallium-208 could serve as a real-time indicator of astrophysical heavy element production, with both rapid ($r$) and intermediate ($i$) neutron capture processes capable of its synthesis. We consider the $r$ process in a Galactic neutron star merger and show Tl-208 to be detectable from $\sim$12 hours to $\sim$10 days, and again $\sim$1$-$20 years post-event. Detection of Tl-208 represents the only identified prospect for a direct signal of lead production (implying gold synthesis), arguing for the importance of future MeV telescope missions which aim to detect Galactic events but may also be able to reach some nearby galaxies in the Local Group.
\end{abstract}

\maketitle

\section{Introduction} 
With LIGO set to observe more neutron star merger events \cite{LIGOO4prospects}, and the prospect of Cosmic Explorer amplifying the number of observations in the future \cite{CosmicExplorer}, finding means to identify individual isotopes from events has become more pressing. Kilonova light curves could point to some individual nuclei if they strongly dominate heating, e.g. Cf-254 \cite{Cfpaper}. Ideally isotope identification could be performed via spectroscopy, and although teasing out absorption features from complex ejecta has provided some insights, this approach has also proven to be challenging \cite{WatsonNature,Nanae21,Vierira23,GW170817TeJWST}. However, the emission spectra of MeV gamma rays can provide a wealth of information when lines can be conclusively attached to a given isotope. In this letter we consider MeV gamma rays emanating from nuclear $\beta$-decays of freshly synthesized isotopes, and highlight for the first time the unmistakable fingerprint of thallium-208 from the $2.6$ MeV line in its emission spectrum.

The $2.6$ MeV gamma ray from thallium-208 is well-known in other fields of science. Clinical imaging studies using the $\alpha$-emitter Ra-224 have found its decay product Tl-208 to be the major contributor to the energy spectra \cite{medimaging2022}. Tl-208 has also been used in nuclear safeguard measures to detect anomalies in nuclear waste since excess count rates for the $2.6$ MeV gamma ray serve as an indicator of shielded highly enriched uranium-232 \cite{wastemat2009}. Due to its production via the Th-232 decay series, thallium's $2.6$ MeV gamma ray is also a well-established background in experimental set-ups, such as SNO \cite{SNO1999} and the neutrinoless double beta-decay experiment Majorana Demonstrator \cite{MajoranaExp}. Additionally, in geology Tl-208 serves to estimate thorium concentrations in aerial surveys \cite{geosurveyIAEAproceeding}, as well as studies of soil \cite{watersoil1971} and snow water content \cite{snowwatercontent1980}. Besides terrestrial uses, the Mars Odyssey gamma-ray spectrometer used the Tl-208 line to infer the presence of Th-232 \cite{MarsOdyssey2006}. The only other mention of Tl-208 in the context of astrophysics is a tentative identification of this species in the COMPTEL instrument line background over 20 years ago \cite{COMPTELbkg2001}. Despite the recognizable nature of Tl-208’s emission line, it has never before been acknowledged that this spectral feature could play a special role in isotope identification for a real-time astrophysical event. Although investigations into MeV gamma-ray emission from neutron star mergers and remnants have been performed \cite{OlegGammas,WuRemnants,ChenGammas,TeradaGammas,Hotokezaka2016}, many focused on longer timescales and none identified the impact of Tl-208 in a real-time event. Additionally, previous studies of real-time signals focused on gamma rays above $\sim$3.5 MeV since these were shown to be exclusively produced when fissioning isotopes are present \cite{fissiongammas}. We highlight that the treatment of decays in the neutron-rich, $A>208$ region (e.g. theoretical $\alpha$- and $\beta$-decay, as well as theoretical neutron-induced, $\beta$-delayed, and spontaneous fission rates) is presently uncertain and could affect the production of Tl-208 on observable timescales.

\begin{figure*}[!t]
    \centering
    \includegraphics[width=8.75cm]{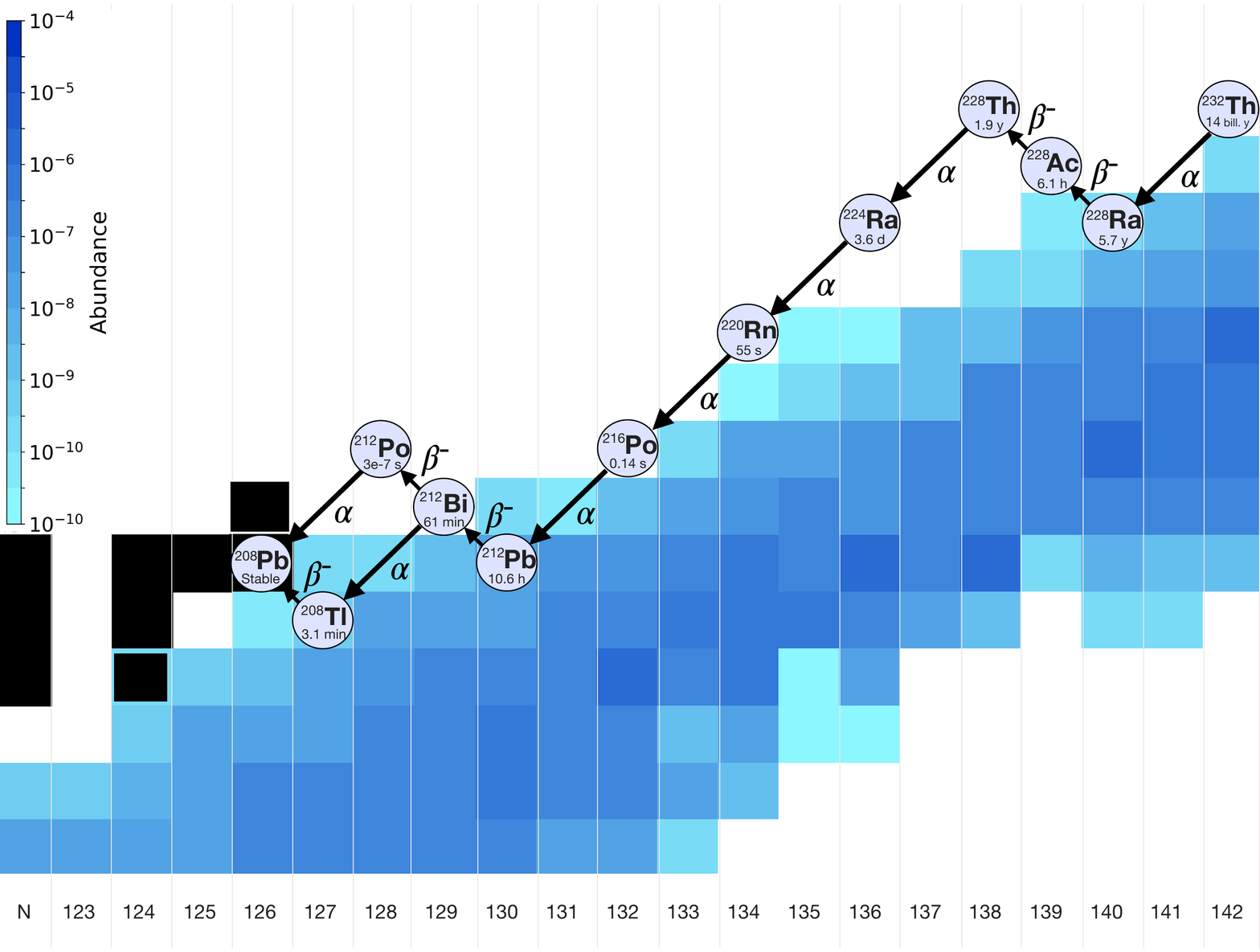}%
    \hspace{0.25cm}
    \includegraphics[width=8.75cm]{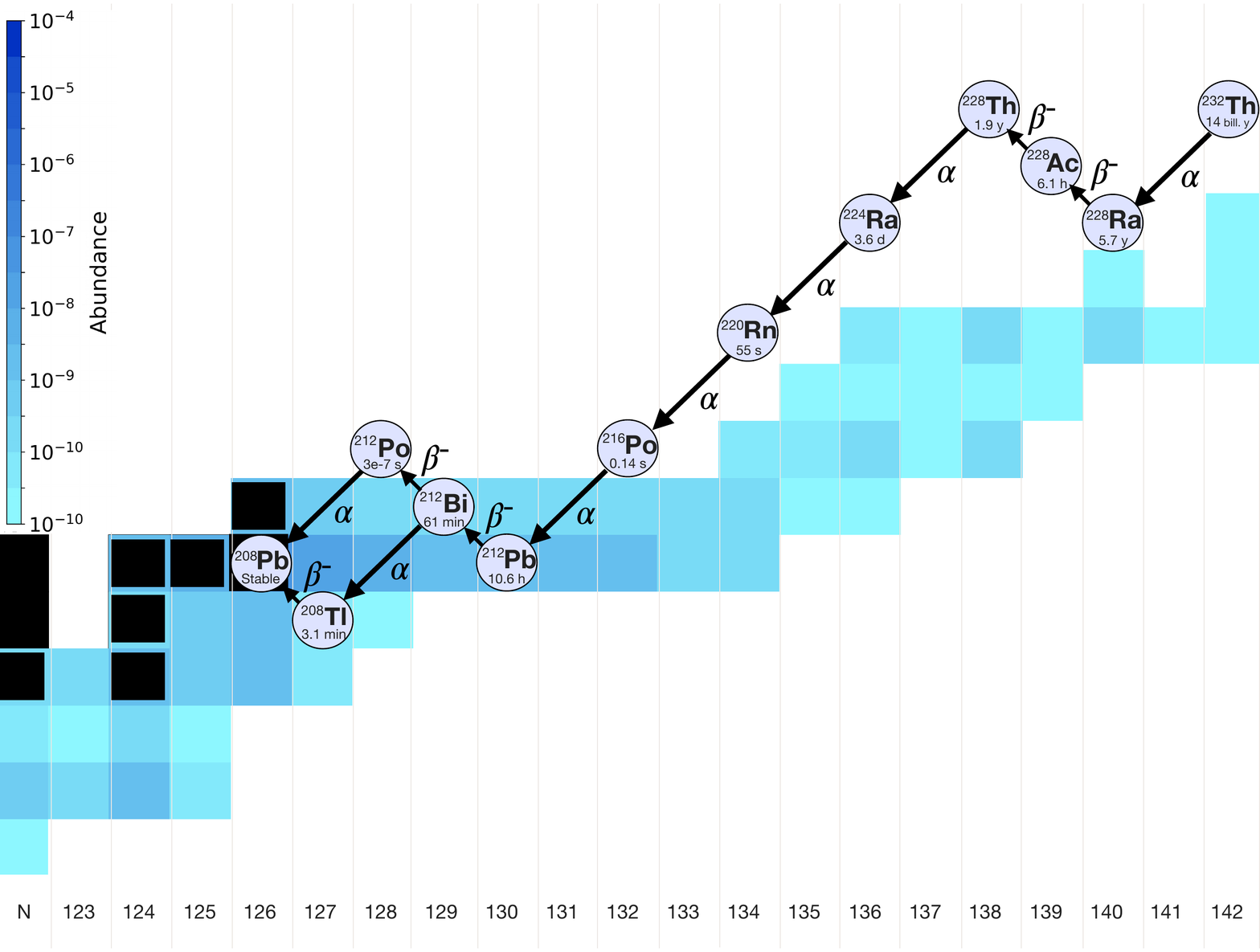}%
   \caption{(Left) Snapshot of $r$-process abundances for a merger ejecta simulation at 6.7 s post-merger during which the decaying nuclei approach species along the Th-232 decay chain. (Right) Snapshot of $i$-process abundances 383 minutes after building from iron seeds in a one-zone model with a neutron density reaching $5\times10^{15}$ cm$^{-3}$.}
\label{fig:abvschain}%
\end{figure*}

In this letter we first discuss the production of Tl-208 in neutron-rich neutron capture nucleosynthesis (i.e. the rapid ($r$) and intermediate ($i$) neutron capture processes). We then focus on the $r$ process in neutron star mergers and show the strong signal that Tl-208 produces on observable timescales and demonstrate that it could be detectable. We lastly explore other gamma rays in the $2.5-2.8$ MeV range, consider thallium identification in the presence of a weak $r$-process component, and provide concluding remarks.

\section{Producing T\MakeLowercase{l}-208 in neutron-rich nucleosynthesis}

Being the direct $\beta$-feeder to lead-208, thallium-208 lies just three neutron numbers outside stability in the neutron-rich regions. Therefore this species is found in a regime reachable by any neutron capture nucleosynthesis process operating outside of stability such as the $r$ process and $i$ process, as is demonstrated in Fig.~\ref{fig:abvschain}. Here the $r$-process scenario corresponds to dynamical ejecta from a $1.2-1.4 M_\odot$ neutron star merger simulation \cite{Rosswog} with $Y_e=\sum_i Z_i Y_i \sim 0.01$ (lower $Y_e$ implies higher neutron-richness). The $i$-process one-zone model considered here is described in \cite{Dardelet,Pavelref2,denissenkov:21} (the $20\%$ H, $50\%$ C case) which represents a scenario in which convective pulses in the He intershell introduce proton ingestion from the hydrogen envelope down to the C-O core allowing for the production of neutrons from $^{13}$C($\alpha$,n) and capable of reaching a neutron density of $\sim5\times10^{15}$ cm$^{-3}$. The astrophysical site(s) for the $i$ process are still unclear. Candidates include post-AGB stars \cite{herwig:11}, low-mass, low-metallicity AGB stars \cite{iwamoto:04,ChoplinAGB}, super-AGB stars \cite{jones:15}, massive stars \cite{clarkson:18,banerjee:18} or rapidly accreting white dwarfs (RAWDs) \cite{denissenkov:17,PavelRAWD}. If $i$-process neutron densities can be maintained for long enough in a stellar environment to reach high enough neutron exposures, actinides can be produced as shown in Fig.~\ref{fig:abvschain} (see also \cite{Chopliniprocactinides}). The total amount of heavy elements produced is uncertain and depends on convective-reactive conditions within the $i$-process site. Here both calculations utilize the PRISM nucleosynthesis network \cite{BDFrp,TrevorDFT} with the nuclear data described in \cite{VasshGEF2019,fissiongammas} (FRDM2012 mass model \cite{FRDM2012}, FRLDM fission barrier heights \cite{FRLDM}, and GEF2016 fission yields \cite{GEF}, with AME2020 \cite{AME2020} and NUBASE2020 \cite{NUBASE2020} experimental data). For experimental gamma-ray spectra, we utilize the ENDF/B-VIII.0 database \cite{ENDF8}.

We first discuss the routes for Tl-208 to be populated within neutron capture nucleosynthesis. One avenue is direct production through the $\beta$-decay of Hg-208, fed in turn by the rest of the $A=208$ decay chain. With Hg-208 having a half-life of 41 min, a thallium gamma-ray signature associated with the $A=208$ chain could take place very early ($\sim$one hour) after an event. Note that $A=208$ half-lives beyond Hg-208 are currently unmeasured, so the approach to Tl-208 via this chain should be reassessed upon future measurements.

More importantly, neutron capture processes can populate species along the well-known Th-232 decay chain, eventually yielding Tl-208. For real-time events, it is not the decay of the long-lived (14 billion year half-life) Th-232 itself which produces Tl-208. Rather neutron capture nucleosynthesis can populate Bi-212, Ra-224, and Ra-228 directly, corresponding to three distinct production timescales for Tl-208.

$Bi$-$212-$ This species dominantly $\beta$-decays to Po-212 which then $\alpha$-decays to Pb-208. However, $\sim$36\% of the time Bi-212 $\alpha$-decays to populate Tl-208 in $\sim$61 min. Bi-212 itself can be populated by the $\beta$-decay of Pb-212 (half-life $\sim$11 h), with the rest of the $A=212$ $\beta$-chain having half-lives of hundreds of ns or less (predicted, not yet measured). Therefore, production of Tl-208 via $A=212$ nuclei corresponds to an emission timescale of $\sim$12 hours. 

$Ra$-$224-$ The $A=224$ chain of Ra-224, Fr-224, Rn-224, and At-224 decay with half-lives of 3.6 days, 3.3 minutes, 107 minutes, and 1.3 minutes, respectively, and no $A=224$ measurements beyond these. Thus Ra-224 decay could provide a timely gamma-ray signal with its production of Tl-208 able to be registered around $\sim$4 days post-event. 

$Ra$-$228-$ Lastly, Ra-228 (5.75 year half-life) is $\beta$ fed by Fr-228 which decays in 38 s. Fr-228 is fed by Rn-228 which $\beta$-decays in 65 sec with the rest of the half-lives along the $A=228$ chain being unmeasured. Ra-228 goes on to $\beta$-decay into Ac-228 which decays in $\sim6$ hours to Th-228 ($\alpha$-decaying to Ra-224 with half-life 1.9 years). Thus Ra-228 ultimately produces Tl-208 via the Ra-224 decay series, but following a several year delay. Therefore Tl-208 gamma-ray emission would be a beacon marking the nucleosynthesis reach up to $A=228$ if observed $\sim$8 years post-event.

As seen in Fig.~\ref{fig:abvschain}, Tl-208 could also be produced in the $i$ process via the already discussed $A=208$, $A=212$, $A=224$, or $A=228$ decay pathways. In addition, here direct neutron capture along the Tl isotopic chain (starting from stable Tl-205) could also produce Tl-208. We find the neutron capture rates for the Tl and Hg isotopic chains of relevance to Tl-208 production (at the 300 MK temperature appropriate in RAWDs) to be uncertain by up to roughly an order of magnitude (using the methodology outlined in \cite{Pavelref2}). With the relevant local $\beta$-decay rates being on the order of minutes or longer, such capture rate uncertainties are not expected to imply Tl-208 production could be foregone entirely, however it should be noted that its abundance could be correspondingly enhanced or reduced relative to the value given in Fig.1.

In $i$-process sites heavy elements will be produced within the He shell at the location of convective pulses and dredge up, thus commenting on the observability of the Tl-208 $2.6$ MeV emission line would require a proper transport calculation, which we leave to future work. Note that $i$-process conditions have been shown based on 3D simulations to be potentially accompanied by a global and possibly violent convective-reactive instabilities \cite{herwig:14} and these may lead to rapid mass ejections \cite{jones:15}. Also, in 1996, during a five month period of the so-called ``very last thermal pulse" of the He shell in a post-AGB star Sakurai's object (V4334 Sagittarii), products of $i$-process nucleosynthesis activated at the neutron density of $\sim 10^{15}$ cm$^{-3}$ were observed to be mixed to the surface the white dwarf \cite{Pavelref1,Pavelref2}. If Tl-208 gammas can pierce through the dynamically ejected He shell, the convective thermal pulses in a RAWD may produce Tl-208 gamma rays with a periodic, repeated signal. $i$-process investigations have indeed suggested that ejecta from environments like RAWDs could contribute to the solar system abundance distribution \cite{Benoitiprocess}. Should such $i$-process ejecta contain actinides, then a real-time signal on the order of days or years is possible. Telescopes have already been used to identify accreting white dwarfs \cite{accretingWDreview} and search for RAWDs in nearby galaxies \cite{searchforRAWD}. Such efforts could provide candidates for future MeV telescope missions to hunt for $in\,situ$ heavy element nucleosynthesis via the Tl-208 beacon.

\section{The T\MakeLowercase{l}-208 emission line from neutron star mergers and detectability}

The $r$ process has long been discussed as a source of the actinides and therefore is clearly capable of producing Tl-208. The kilonova observation for the neutron star merger event GW170817 confirmed lanthanide ($57<Z<71$ with stable isotopes at $139<A<176$) production \cite{Cowperthwaite2017}, but could not yield definitive statements regarding elements beyond this. Although the effect of Cf-254 on kilonova light curves could point to actinide production \cite{Cfpaper}, specific indicators of elements between Cf and the lanthanides, such as lead, have not been previously highlighted. Here we demonstrate that Tl-208 could provide a smoking gun of lead and third $r$-process peak (at $A\sim195$ containing elements like gold and platinum) production. Tl-208 gamma-ray emission can even indicate a minimum $r$-process reach, pointing to at least $A=224$ or $228$ if detected on the order of days or years post-merger, respectively. 

\begin{figure}
    \centering
    \includegraphics[width=7.75cm]{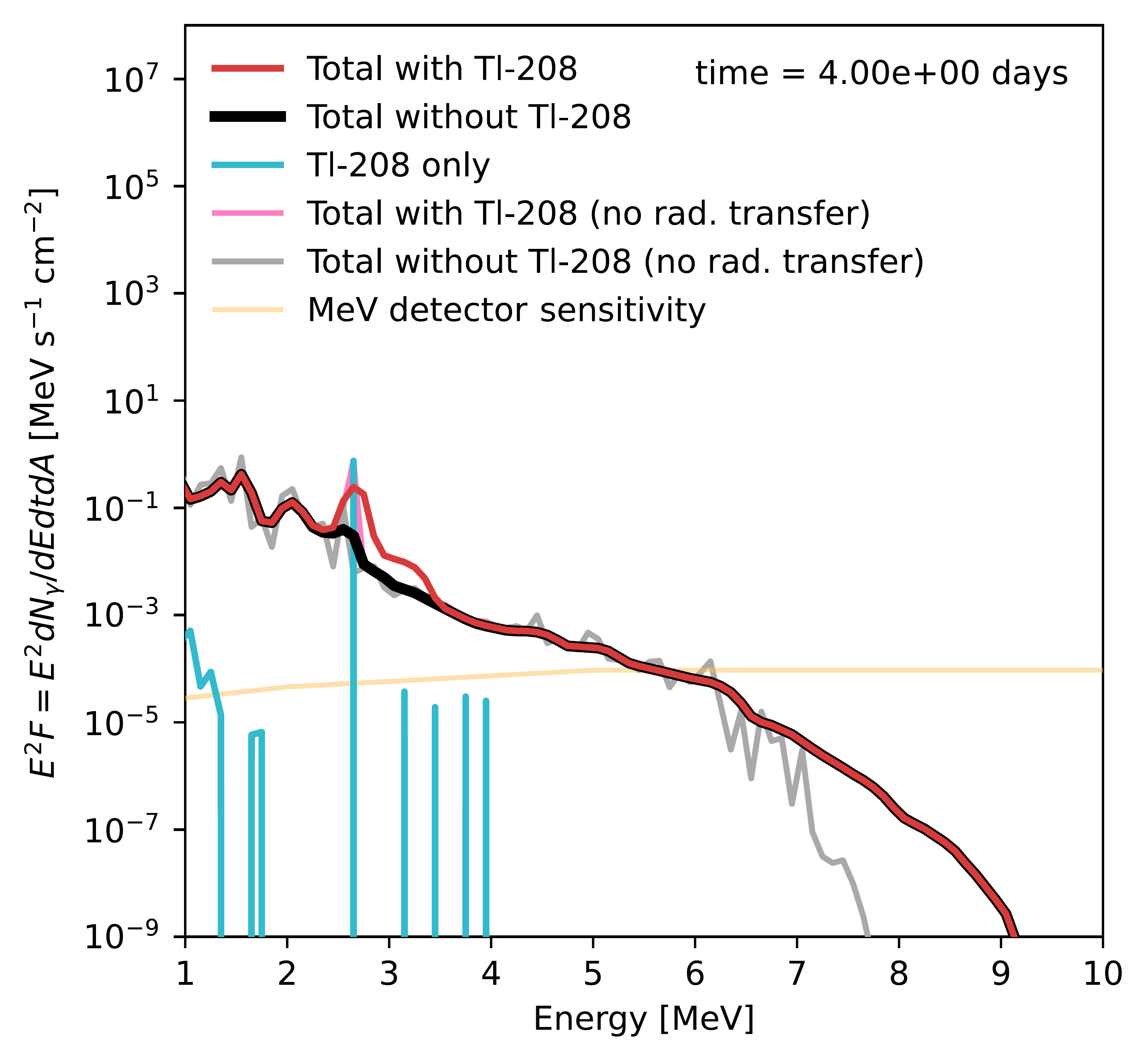}%
    \hspace{0.125cm}
    \includegraphics[width=7.75cm]{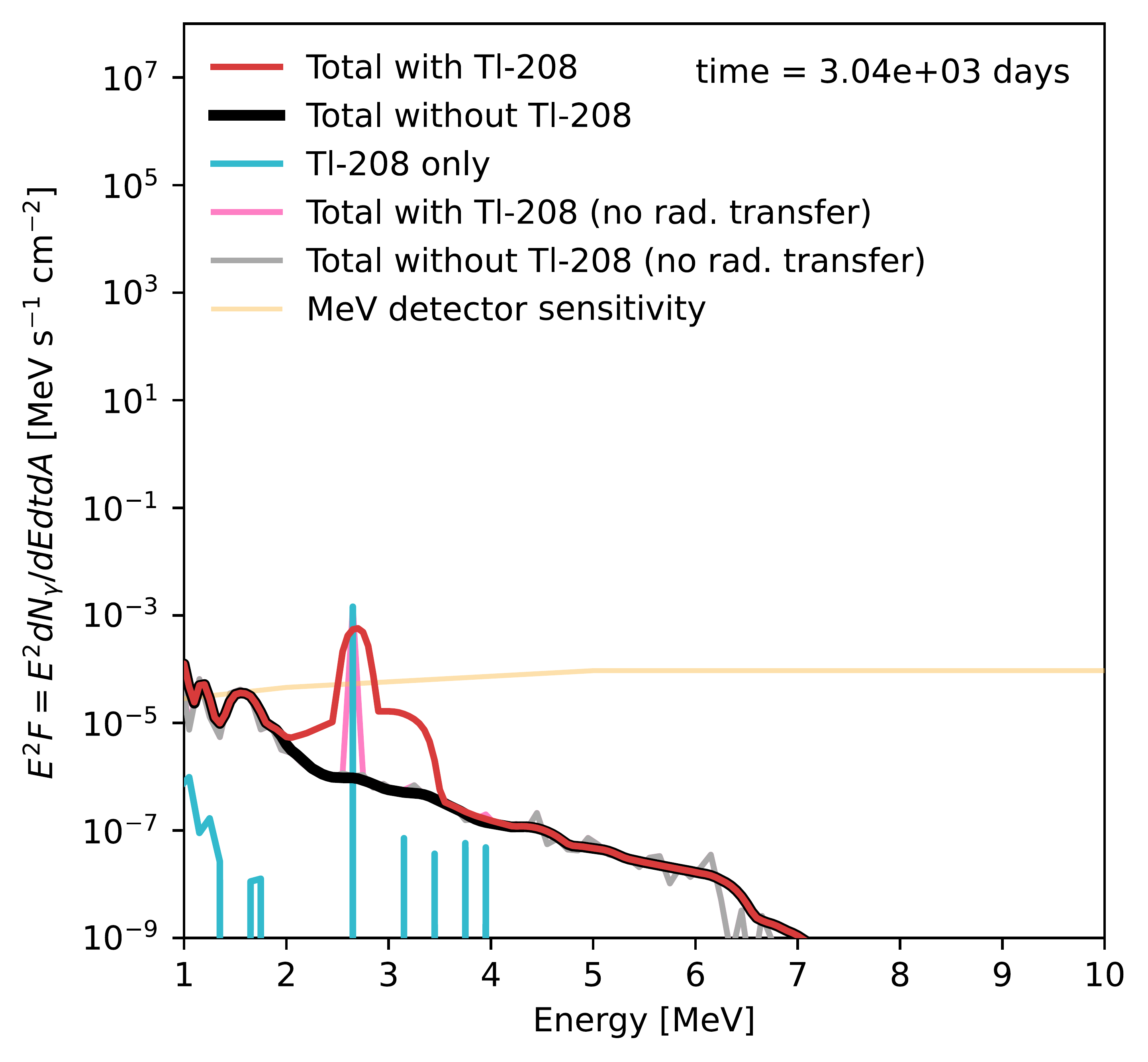}%
    \caption{The total $\beta$-decay gamma-ray spectrum from a Galactic neutron star merger at 10 kpc compared to the Tl-208 spectrum alone and the total without Tl-208 during the ~12 hours to 10 days period (top) after Tl is populated by Bi-212 and Ra-224 decay, followed by emission ~1-20 years post-event (bottom) due to Ra-228 decay. Also shown is the estimated sensitivity limit for next generation MeV detectors like MeVGRO with $\sim1$ day exposure time.}
\label{fig:tlonlynotlspectra}%
\end{figure}

In figure~\ref{fig:tlonlynotlspectra} we see the $2.6$ MeV Tl-208 line explicitly show itself in the spectrum at the distinct timescales mentioned. Here we consider a Galactic merger at 10 kpc and propagate the gamma-rays emitted from $\beta$-decays through the merger ejecta using the same radiation transfer methods outlined in \cite{fissiongammas}, where we assume uniform, spherical $r$-process ejecta expanding homologously with the ejecta consisting of an inner core plus an outer shell and a velocity at the outermost radius of 0.3c (note that a lower velocity would give more distinctive lines which are less Doppler broadened). The propagating gamma-ray photon interactions include Rayleigh scattering, Compton scattering, photoelectric absorption, and pair production. The opacity or cross-section for the photon interactions with ejecta are calculated based on the ejecta's composition with a mixture of the opacities of characteristic isotopes. On timescales of days and years, the $\beta$-decays back to stability of $r$-process species with higher Q-values have mostly ceased, allowing the $2.6$ MeV line to be clearly distinguished. Thus Tl-208 can serve to decipher emission from a merger event at more than one timescale.

Importantly, for an event that occurs within our Galaxy, the gamma-ray spectrum for all possible timescales of Tl-208 detection is above the estimated sensitivity limit with $\sim1$ day exposure time of proposed next generation MeV telescopes like MeVGRO \footnote{\url{https://indico.icranet.org/event/1/contributions/777/}}. The earlier Tl-208 signal (on the order of days) can also be observed by COSI \footnote{\url{https://cosi.ssl.berkeley.edu}}, which will be launched in a few years. Additionally, we find that this earlier signal could be observable out to $\sim$200 kpc (from increasing the distance assumed until the predicted signal meets detector sensitivity). Therefore Tl-208 emission may even be visible from extragalactic events should they occur in nearby Local Group galaxies, such as the SMC and LMC, within hundreds of kpc.

\section{Observability of T\MakeLowercase{l}-208 over emission from other nuclei}

\begin{figure}[!b]
    \centering
    \includegraphics[width=8.7cm]{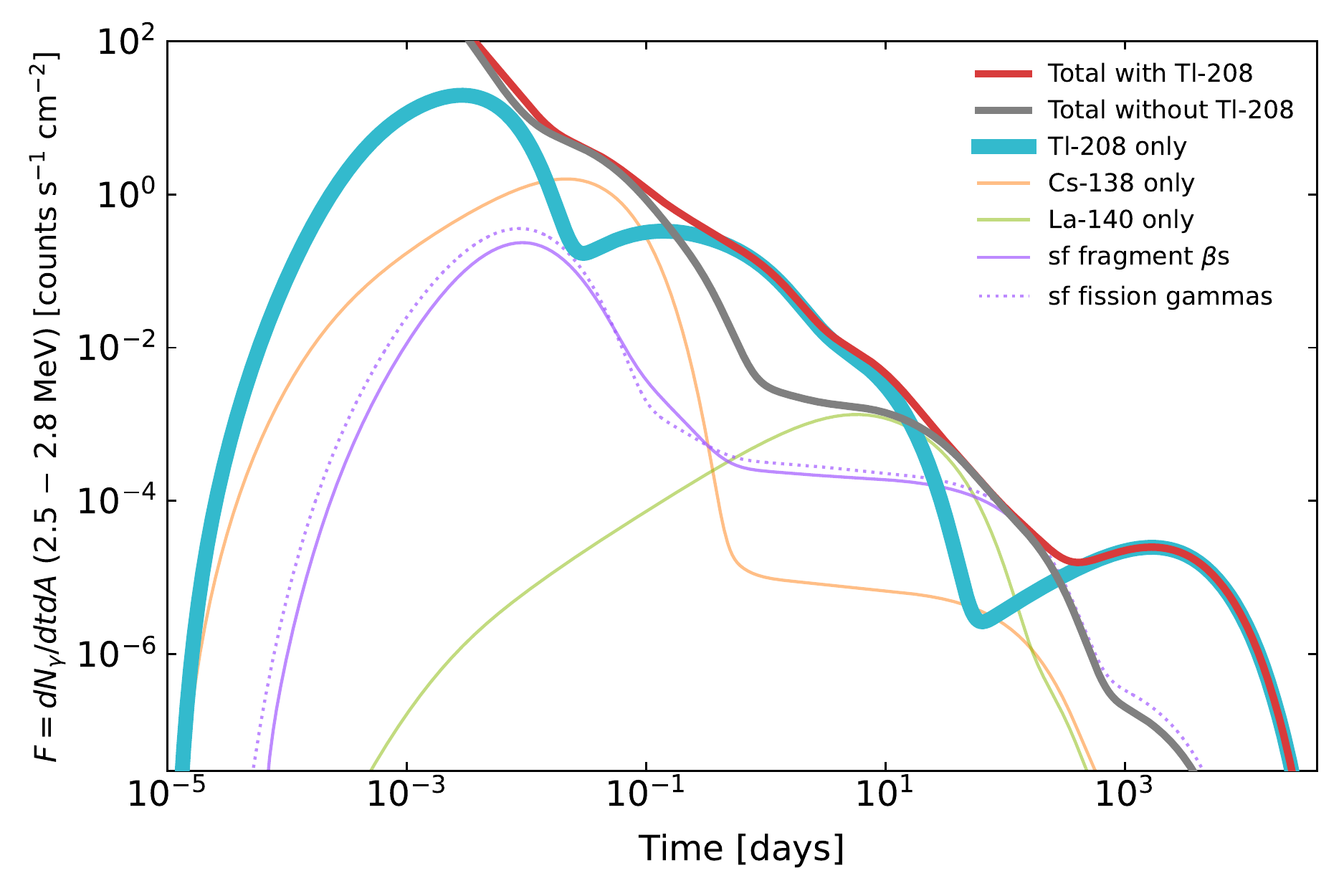}%
  \caption{The light curve (flux integrated) over the $2.5\le E_{\gamma} \le2.8$ MeV range with and without Tl-208 as compared to the contribution from Tl-208 alone (all before radiation transfer). Also shown are other species whose $\beta$-decay gammas dominate at these energies, as well as the prompt gammas from Cf-254 spontaneous fission (sf) and the subsequent gammas from the $\beta$-decay of its fragments.} 
\label{fig:summedgam}%
\end{figure}

We next evaluate the observability of Tl-208 over other species emitting in the $2.5-2.8$ MeV range (surrounding the key $2.6$ MeV spectral line). Figure~\ref{fig:summedgam} shows the case where the merger is dominated by dynamical ejecta (with mass ejection of $0.01 M_\odot$) and compares the total ``emitted” light curve (before radiation transfer) with and without Tl-208 alongside the individual contribution from thallium. We see a clear dominance of Tl-208 at $\sim$12 hours to a few days (from $A=212$ and $A=224$ nuclei respectively), followed by a strong peak a few years post-event from $A=228$ nuclei, with the light curve having a fundamentally different evolution in the absence of Tl emission. At the distinct emission timescales of Tl-208 we do not find significant competition from the decay spectra of other nuclei. Emission from Sb-131 and Cs-138 (and to a lesser degree La-142 and Sn-127) instead dominates just prior to the emergence of Tl emission on the order of days. Later, during in the window between the Tl-208 emission times of days and years, we see emission from La-140 as well as the prompt fission gammas from Cf-254 alongside the $\beta$-decay gammas from its neutron-rich daughters. Since our previous work on fission gammas \cite{fissiongammas} only explored the energies above 3 MeV, Fig.~\ref{fig:summedgam} represents the first evidence that fission gammas compete with gamma rays from $\beta$-decay in the $2.5-3$ MeV regime, and could be observable around $\sim$100 days. Note the adopted fission gamma-ray spectra are from theoretical predictions (described in \cite{fissiongammas}) and therefore the predicted role of fission gammas in this energy regime could change in the future. Additionally, we note that recent studies have shown that isomers may play a role in setting decay timescales \cite{Wendell}, but are unlikely to impact this study given the well-known nature of the Th-232 decay chain, as well as the distinct nature of Tl-208’s strong 2.6 MeV line. Ultimately, we find that no other nuclei produced in dynamical ejecta with emission lines in the $2.5-2.8$ MeV range obscure post-merger Tl-208 signals.

\begin{figure}[!h]
    \centering
    \includegraphics[width=8.7cm]{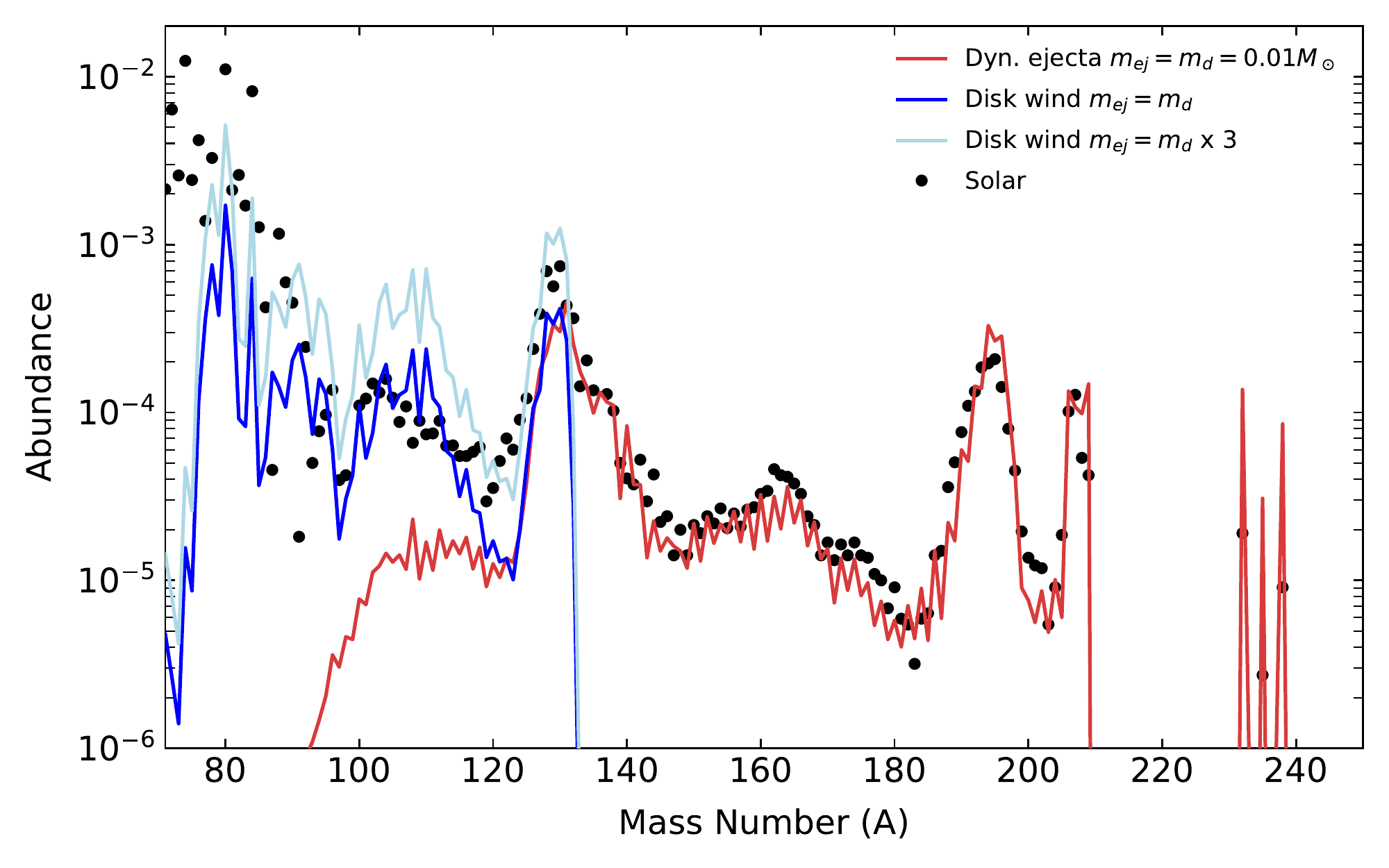}%
    \hspace{0.125cm}
    \includegraphics[width=8.7cm]{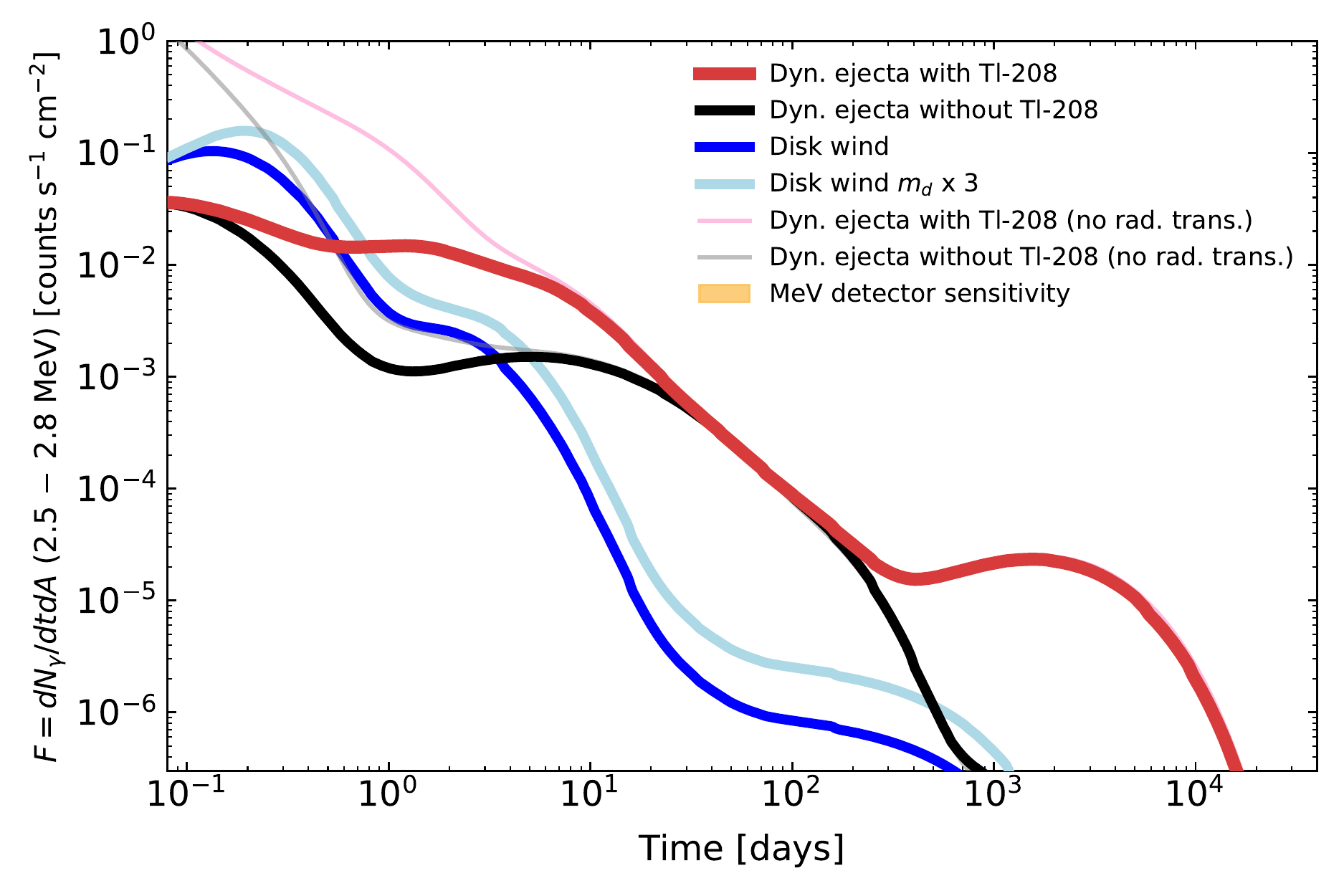}%
   \caption{(Top) Abundance patterns (compared to Solar \cite{Sneden}) for merger dynamical ejecta \cite{Rosswog} with $Y_e\sim0.01$ and a parameterized disk wind (entropy of $s=30$, dynamical timescale of $\tau=70$ ms, with $Y_e=0.3$). (Bottom) Comparison of the light curve (in the $2.5\le E_{\gamma} \le2.8$ MeV range) for the disk versus the dynamical (with Tl-208 and without Tl-208), along with the MeV detector sensitivity.}
\label{fig:weakvstrong}%
\end{figure}

Finally, since the presence of a weak (lanthanide-free) component in mergers, for example from an accretion disk wind, can be inferred from the early blue kilonova emission of GW170817 \cite{Cowperthwaite2017}, we consider the previous dynamical ejecta case alongside a weak r process. Figure~\ref{fig:weakvstrong} demonstrates that a weak component could emit in the $2.5-2.8$ MeV range, particularly dominated by Ag-112 and Rh-106. Importantly, even when considering a weak component with three times the mass ejection of the dynamical component, emission from Ag-112 and Rh-106 are subdominant to the $2.6$ MeV Tl-208 gamma-ray signature. Thus on the emission timescales of Tl-208, this species consistently outshines all other nuclei producing emission lines in the $2.5-2.8$ MeV window.

\section{Conclusions}

In this letter we have shown that the well-known fingerprint of Tl-208's $2.6$ MeV gamma-ray line, utilized in numerous other branches of science, serves an important role in astrophysics as a distinct indicator of heavy element production in real-time neutron capture nucleosynthesis. We demonstrated its ability to be produced in both the astrophysical $r$ process and $i$ process since, being the direct $\beta$-feeder to Pb-208, Tl-208 does not lie far outside stability. With Tl-208 being close to stability and connected to nearby long-lived actinides, clear opportunities exist for future measurements at ARIEL at TRIUMF, the N=126 Factory at ANL, and FRIB to pin down the local capture and decay information impacting predictions for Tl-208 gamma-ray emission. 

Importantly, via Tl-208's place as an endpoint on the Th-232 decay chain, we have shown for the first time that signals from the $2.6$ MeV thallium line could be observable by next generation MeV telescopes during several time windows following a nearby neutron star merger. Tl-208 serves as a smoking gun of the reach of an $in\,situ$ neutron capture nucleosynthesis event, indicating the production of at least Pb-212 if seen $\sim$12 hours post-event, Ra-224 if the signal continues for several days, and Ra-228 should Tl-208 be observed a few years post-event for a merger within our Galaxy. The signal at days or earlier could be observed not only for Galactic events, but out to Local Group galaxies such as the LMC and SMC.

For detectability, it is important to note that $r$-process events are known to be rare, but their rate is uncertain, with the rate in the Milky Way from a recent study based on merger estimates of known Galactic neutron star binaries reported to be $25^{+19}_{-9}$ Myr$^{-1}$ \cite{ColomiBernadich} and another study which made use of the ratio of local merger rates to core-collapse supernova rates finding a Milky Way merger rate of $\sim440$ Myr$^{-1}$ \cite{XiluNESN}. Note that since other galaxies such as the LMC and SMC are reachable and also have active star formation, the Milky Way merger rate should be considered a lower bound on the possible event rate for a detectable $r$-process Tl-208 signal. Regarding detectability from an $i$-process event, here the rate is very uncertain since the possible site(s) are still under active investigation.  Low-mass AGB stars and super-AGB stars at very low metallicities could host the $i$ process \cite{superAGB}, and such stars can be found in nearby globular clusters with [Fe/H]$<-$2 such as M92, M15, and M30 at distances of $\sim$5-10 kpc. In the case of RAWDs, the $i$ process could occur at any metallicity (including Solar) but the maximum neutron density decreases with an increasing metallicity, and so it may be difficult for RAWD nucleosynthesis to reach thallium at some metallicities such as Solar. The birth rate of RAWDs has been estimated for the present day to be between 500-700 Myr$^{-1}$ \cite{Benoitiprocess} in the Milky Way galaxy. Therefore, since $i$-process candidate AGB sites can be found within the Milky Way and the rate of RAWDs could be comparable to or greater than the local neutron star merger rate, searching locally for $in\ situ$ $i$-process heavy element production through Tl-208 emission is possible.

The identification of the thallium line represents the only known current direct signal of lead production (implying gold synthesis) in a real-time event. Thus this clear, distinct signature of real-time heavy element production provides a key opportunity for future MeV gamma ray missions, such as MeVGRO and AMEGO\footnote{\url{https://asd.gsfc.nasa.gov/amego/}}, to map out a detailed picture of the isotope composition of events, leading to more definitive statements regarding the ultimate role each event type plays in the astrophysical origin of elements.

\section{Acknowledgments}
N.V. would like to thank Benoit C{\^o}t{\'e} for useful discussions. N.V. and M.L. acknowledge the support of the Natural Sciences and Engineering Research Council of Canada (NSERC). 
The work of X.W. was supported by National Key R\&D Program of China (2021YFA0718500) and the Chinese Academy of Sciences (Grant No. E329A6M1). 
The work of N.V., G.C.M., M.R.M., and R.S. was partly supported by the Fission In R-process Elements (FIRE) topical collaboration in nuclear theory, funded by the U.S. Department of Energy. Additional support was provided by the U.S. Department of Energy through contract numbers DE-FG02-02ER41216 (G.C.M), DE-FG02-95-ER40934 (R.S.), and LA22-ML-DE-FOA-2440 (G.C.M. and R.S.). X.W., R.S. and G.C.M also acknowledge support by the National Science Foundation Grants No. PHY-1630782 and PHY-2020275 (Network for Neutrinos, Nuclear Astrophysics and Symmetries). M.R.M. was supported by the US Department of Energy through the Los Alamos National Laboratory. Los Alamos National Laboratory is operated by Triad National Security, LLC, for the National Nuclear Security Administration of U.S.\ Department of Energy (Contract No.\ 89233218CNA000001). Zhenghai Liu was supported by a contract from Los Alamos National Laboratory. G. C. M. and R.S. acknowledge support through Exascale Nuclear Astrophysics for FRIB (ENAF) a DOE SciDac project.  FH acknowledges funding through an NSERC Discovery Grant. PD acknowledges CaNPAN support through NSERC under award SAPPJ-2021-00032. This manuscript has been released via Los Alamos National Laboratory report number LA-UR-23-27995.

\bibliography{Tlrefs}
\bibliographystyle{apsrev4-1}

\end{document}